\documentclass{article}
\usepackage[margin=3cm]{geometry}
\usepackage{amsmath}
\usepackage[latin1]{inputenc}
\usepackage{graphicx}
\usepackage{tikz}
\usepgflibrary{snakes}
\DeclareGraphicsExtensions{.jpg,.pdf,.png}

\renewcommand{\v}[1]{\boldsymbol{#1}}

\title{Synchronization and control of cellular automata}
\author{Franco Bagnoli$^{(1)}$ and Ra\'ul Rechtman$^{(2)}$\\
 (1) Dipartimento di Energetica, \\Universit\`a di Firenze, Firenze, Italy; also CSDC and INFN, sez. Firenze.\\franco.bagnoli@unifi.it\\
 (2)  Centro de Investigaci\'on en Energ\'\i a, \\Universidad Nacional Aut\'onoma de M\'exico, Temixco, Mor., Mexico\\
    rrs@cie.unam.mx
}
\begin{document}

\maketitle

\begin{abstract}
The control of chaotic systems implies inducing an unpredictable system to follow a desired trajectory using the smallest ``force''. In low-dimensional 
continuous systems, one method is that of reconstructing the tangent space, 
so that the control may be concentrated on the most expanding directions. 
This scheme
is hard to follow for high-dimensional (extended) systems. This is
particularly true for extended systems that exhibit {\em stable chaos},
that is, systems which are not chaotic in the usual sense, but are 
unpredictable for finite perturbations. Prototypical of this class are cellular automata, aka completely discrete dynamical systems. Although usual indicators of chaoticity such as the maximum Lyapunov exponent may be defined for such systems, we show that the usual approach may lead to counter intuitive results, and that it is possible to exploit the characteristics of the system in order to reduce the distance between two replicas with less control.
\end{abstract}

\section{Introduction}
Control theory is a set of techniques for making a dynamical system 
behave in a desired way exerting a minimum effort. In particular, this 
technique may be applied to chaotic systems in order to make them follow a 
desired periodic orbit~\cite{Ott} or to synchronize a ``slave'' replica 
with a ``master'' one~\cite{ReplicaSynchro}.
Most of the literature about control theory deals with low-dimensional 
systems
modeled by a few differential equations. In such systems, the number of 
expanding directions is small and they behave smoothly in 
tangent space.

In this paper we want to introduce the problem of controlling \emph{extended, highly non-linear} dynamical systems.
There is a class of systems, termed \emph{stable chaotic}~\cite{StableChaos} which are nor chaotic in the usual sense of the sensitive dependence with respect to infinitesimal perturbations, but are nonetheless unpredictable.
In particular, we shall concentrate on cellular automata (CA), which are discrete, deterministic dynamical systems.
CA are widely used  to model many systems in various fields, from computer science to earth sciences, biology, physics, sociology, etc. They are usually defined on a graph or a regular lattice, but may easily be extended to include mobile agents. The modeling of a system using cellular automata is conceptually much simpler than those using partial derivatives, and the evolution of such a system is easily performed by a digital computer, without rounding errors.
However, for such systems continuity and smoothness (differentiability) do 
not apply. It is therefore hard to extend the usual techniques used in 
control thory and to define quantities like Lyapunov exponents and
chaotic trajectories.

In the case of master-slave synchronization, the ``minimal strength'' needed
to synchronize a system is related to its chaoticity, defined by the 
lagest Lyapunov exponent in low-dimensional systems. For extended systems, 
the correspondence between the minimal strength and Lyapunov exponents 
may break down~\cite{MapSynchro}.
It is still possible to define derivatives of discrete 
systems~\cite{BoolDeriv}, which prove useful in 
synchronization investigations~\cite{CASynchro}.

In synchronization experiments, the  ``force'' is generally applied blindly, without any relation with the dynamics. The corresponding synchronization effect is analogous to a directed  percolation phase transition. The two systems synchronize when their difference goes to zero. Their difference grows due to their ``chaotic'' dynamics, along the directions identified by the 
Jacobian matrix of the evolution rule. The synchronization ``pressure'' reduces the paths along which a difference can propagate. When this reduction overcomes the chaotic growth, the system synchronizes. 

In control problems, one wants to exploit the knowledge about a system. It is therefore analogous to a synchronization problem of two different systems with a ``targeted'' force, that tries to ``kill'' the growing directions of the difference as soon as possible. We show how the concept of Boolean derivative and that of Boolean Jacobian matrix can be used to achieve this goal.

This technique may be used also as a measurement technique: one system may be the experimental one, and the replica may be simulated on a computer. In this case, the two systems are in general different. We apply the control technique in order to synchronize two different cellular automata rules, one of which may be stochastic (i.e., influenced by external noise). 
	
\section{Definitions}

Let us start our presentation by considering two smooth, chaotic maps 
\[
  \left\{\begin{aligned}
   x' &= f(x),\\
   y' &= (1-p) g(y) + p f(x),
  \end{aligned}\right.
\]
where $p$ is the control ``strength'', $f$ and $g$ are two maps, $x$ is
valued at the discrete time $t$, and $x'$ is valued at $t+1$ (the same 
applies to $y$ and $y'$). The $x$ is the ``master'', the $y$ the ``slave''.
The separation between both maps is $u=x-y$ and 
the goal of control is to keep $h=|u|$ below a certain threshold (which 
may be zero), using the minimum strength $p$. The two maps $f$ and $g$ may 
be different (for instance, they may use different parameters) or the same, 
in this case (synchronization) the synchronized state $x=y$ is absorbing.
	
If the desired trajectory is a natural one for the slave, control 
is equivalent to master-slave synchronization.
\[
 \left\{
 \begin{aligned}
    x' &= f(x),\\
    y' &= (1-p) f(y) + p f(x),
  \end{aligned}\right.
\]
and 
\[
    u' = (1-p) \bigl(f(x)-f(y)\bigr).
\]
For smooth maps, near the synchronization threshold $p_c$, it is possible to expand $y(t)$ around the unperturbed trajectory $x(t)$,
\[
  u' = (1-p) \bigl(f(x)-f(y)\bigr) \simeq (1-p) \frac{d f(x)}{dx} u.
\]
By iterating this map, one obtains the relation between synchronization threshold $p_c$ and Lyapunov exponent $\lambda$,
\begin{equation}
  p_c = 1-\exp(-\lambda).
\end{equation}
The synchronized state is absorbing, since if for some time $x(t)=y(t)$, then the control can be relaxed and the trajectories stay synchronized. However, for chaotic systems, this state is unstable.

%

\subsection{Extended systems}

Natural systems, however, are rarely low-dimensional. We can extend the previous analysis by considering 
a lattice of coupled maps, that may be thought as a stroboscopic view of a continuous system:
\begin{equation}\label{extended}
 x_i(t+1) = f(g(x_{i-1}(t), x_i(t), x_{i+1}(t))).
 \end{equation}
where $i=1,\dots,N$.
The function $g$ represents the spatial coupling, it can be diffusive (linear) or highly nonlinear. The function $f$ is the individual map, and can lead, when uncoupled, either to fixed points, stable cycles or chaotic oscillations. 
 A perturbation may amplify exponentially in time by the action of $f$,  but only linearly in time through the coupling (propagation to neighboring sites). 

The dynamical properties of an extended system are generally analyzed by means of the Lyapunov spectrum. Using vector notation, Eq.~(\ref{extended})
can be written as
\[
  \v{x}(t+1) = \v{F} (\v{x}(t)).
\]
The components of the Jacobian matrix of $\v{F}$ are 
\[
 J_{ij}(\v{x}(t)) = \dfrac{\partial F_i(\v{x}(t))}{\partial x_j}
\]
$i,j=1,\dots,N$. For an infinitesimal perturbation
\[
  \v{\delta}(t+1) = \v{J}(\v{x}(t)) \v{\delta}(t).
\]
For instance, the Jacobian matrix of one-dimensional systems with nearest 
neighbor couplings has zero values except on the three central 
diagonals as 
\[
    \v{J} = \begin{pmatrix}
      J_{1,1} & J_{1,2} & 0 & 0 & \cdots&0 & J_{1,N}\\
      J_{2,1} & J_{2,2} & J_{2,3} & 0 & \cdots& 0 & 0\\
      0 & J_{3,2} & J_{3,3} & J_{3,4} & \cdots &0 & 0\\
      \vdots & \vdots& \vdots&\vdots&\ddots &&\vdots\\
     \vdots & \vdots& \vdots&\vdots& &\ddots&\vdots\\
       J_{N,1} & 0 & 0 & 0 & \cdots& J_{N-1,N}& J_{N,N}\\
       \end{pmatrix}.
\]
The eigenvectors of the Jacobian define the instantaneous \emph{tangent space} of a dynamical system. The eigenvalues $a_i$ of the time product $\prod_{t=0}^T \v{J}(\v{x}(t))$ of the Jacobian matrices over
a trajectory define the Lyapunov spectrum 
$\lambda_0\ge\lambda_1 \ge\lambda_2\dots$ with $\lambda_i=\log(a_i)/T$~\cite{ott92}. 
It is generally
assumed that a system is chaotic if $\lambda_0>0$ and stable if
$\lambda_0<0$.

The largest Lyapunov exponent $\lambda_0$ (LLE) does not capture all 
the chaotic characteristics of an extended system. In general, a weak 
diffusive coupling reduces the LLE (since diffusion limits the exponential 
expansion along tangent space). Therefore, for small couplings, the maximum 
of chaoticity corresponds to \emph{uncoupled} maps, but this situation
may correspond to the easiest synchronizability (see Section~\ref{sec:synchro}).

\begin{figure}
 \begin{center}
  \begin{tabular}{cc}
    \includegraphics[width=0.385\textwidth]{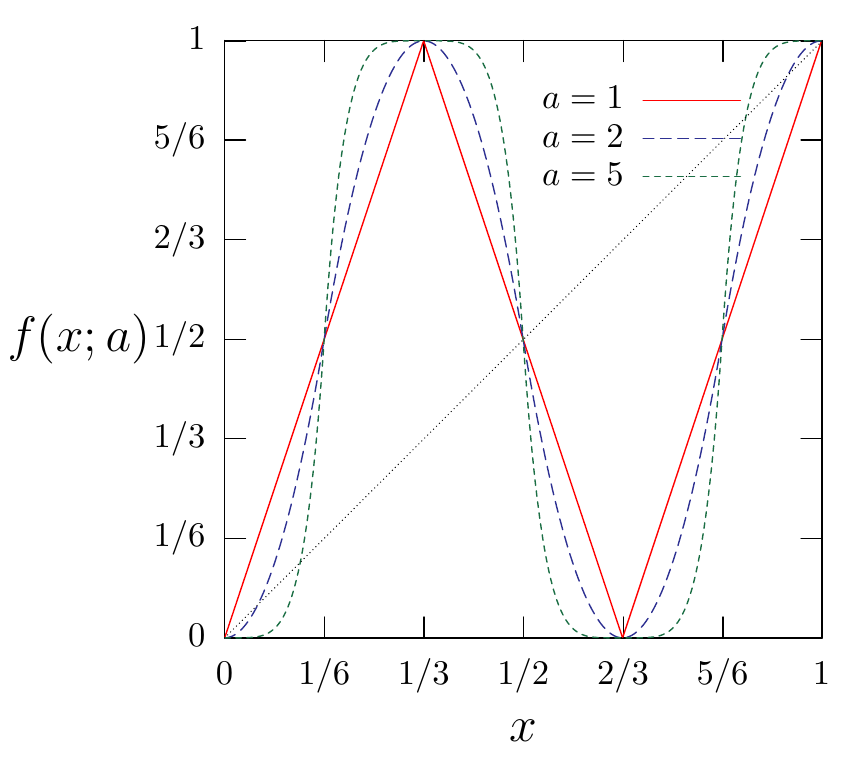} &
    \includegraphics[width=0.65\textwidth,viewport=0 -25 300 100]{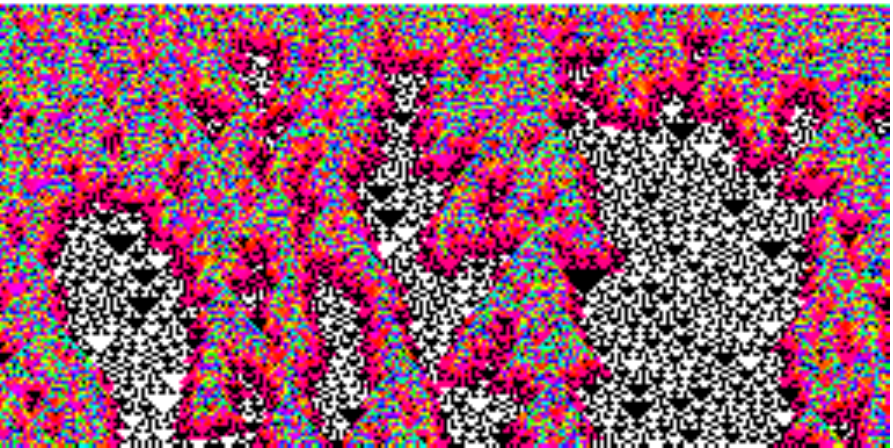}\\
    (a) & (b)~~~~~~~~~~~~~~~\\
  \end{tabular}
 \end{center}
  \caption{\label{fig:map} (a) The plot of the  map of \eqref{map} for different values of the parameter $a$. For $a>1$, this map exhibits two attracting superstable fixed points. (b) Time evolution (downward) of a lattice of coupled maps. Color code: white=0, black=1, color=intermediate values. One can observe a transient disordered evolution (transient chaos) followed by a cellular automata pattern.
  }
\end{figure}

\subsection{Stable chaos and cellular automata}
The scenario of extended systems may be more complex as we discuss below. 
Consider the map defined below and shown in Figure~\ref{fig:map}-a)
\begin{equation}
 \label{map}
 f(x;a)=
 \begin{cases}
  (6x)^a/2    & 0\leq x < 1/6,\\
  1-|6(1/3-x)|^a/2  & 1/3\leq x < 1/2,\\
  |6(x-2/3)|^a/2  & 1/2\leq x < 5/6,\\
  1-(6(1-x))^a/2  & 5/6\leq x < 1,
 \end{cases}        
\end{equation}
This map is obviously stable for $a>1$, with two fixed points $x_0=0$ and $x_1=1$, with interleaved basins that act as a sort of ``frustration'' when coupled,
\begin{equation}
x_i(t+1) = f\left(\frac{x_{i-1}(t)+ x_i(t)+ x_{i+1}(t)}{3} \right),
\end{equation}
$i=1,\dots,N$ with periodic noundary conditions.
The system continues to be stable, exhibiting transient chaos 
(see Figure~\ref{fig:map}-b). After a transient, the eigenvalues of
the Jacobian matrix  go to zero, and all Lyapunov exponents go to $-\infty$.
After a transient, the evolution is that of cellular automatn rule 150 
(see Section~\ref{sec:CA}), which is unpredictable for \emph{finite} 
perturbations greater that $1/6$~\cite{mappet}. A similar behavior can be 
found in other continuous systems without direct correspondence to 
cellular automata~\cite{StableChaos}.
Unpredictable stable systems are interesting since the synchronized state is stable.

\begin{figure}
   \begin{center}
    \begin{tabular}{|ccc|c|c|c|c|c|c|c|c|c|}
     \hline
  $x_0$ & $x_1$ & $x_2$ & $t$ & $f$ 
  & $\partial f/\partial x_0$
  & $\partial f/\partial x_1$
  & $\partial f/\partial x_2$        \\[1mm]
   \hline                                             
  0    & 0 & 0 &0  & 0 & 1 & 1 & 1          \\
  0    & 0 & 1 &1  & 1 & 1 & 1 & 1          \\
  0    & 1 & 0 &1  & 0 & 1 & 1 & 1          \\
  0    & 1 & 1 &2  & 1 & 1 & 1 & 1          \\
  1    & 0 & 0 &1  & 1 & 1 & 1 & 1          \\
  1    & 0 & 1 &2  & 0 & 1 & 1 & 1          \\
  1    & 1 & 0 &3  & 1 & 1 & 1 & 1          \\
  1    & 1 & 1 &3  & 0 & 1 & 1 & 1          \\
  \hline
    \end{tabular}
   \end{center}
  \caption{\label{fig:10} Possible inputs ($x_0, x_1, x_2$), sum ($t$), output ($f$) and first-order derivatives of totalistic function $R=1010|_2=10|_{10}$. Since derivatives do not depend on configurations, it is a linear rule. }
\end{figure}

\begin{figure}[b]
  \begin{tabular}{ccc}
   \includegraphics[width=0.3\textwidth]{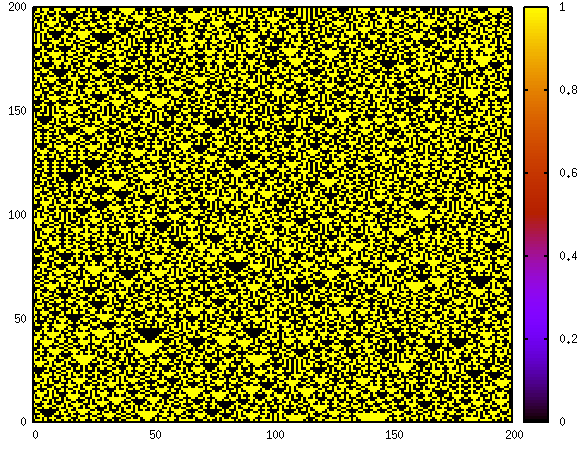}  &
   \includegraphics[width=0.3\textwidth]{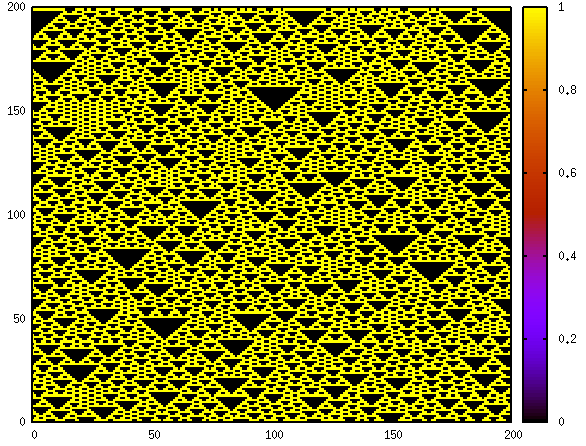}  &
   \includegraphics[width=0.3\textwidth]{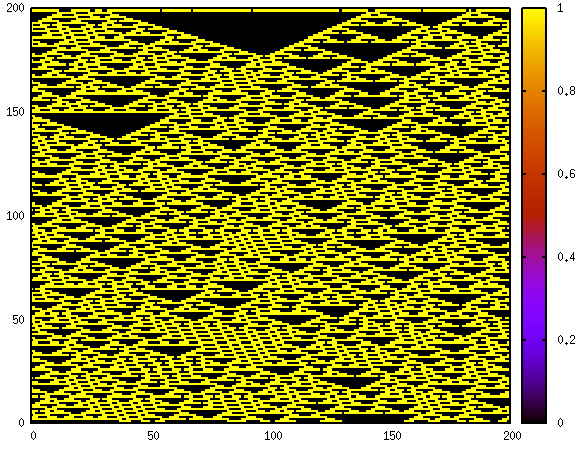}  \\
   $r=3, R=10$&
   $r=3, R=6$&
   $r=6, R=30$
  \end{tabular}
  \caption{\label{fig:pattern} Typical space-time patterns of ``chaotic'' rules.}
\end{figure}

\section{Cellular automata}\label{sec:CA}

Cellular automata (CA) are completely discrete systems, defined as in 
Eq.~\eqref{extended}, where $x_i$ and $f$ can assume values in a discrete
set. In particular, we shall limit our study to Boolean CA, for which the
set of discrete values is $\{0,1\}$.
Since the function $f$ is discrete, it can be defined by means of a complete enumeration of output given all possible inputs (look-up table). We shall denote by $r$ the size of the neighborhood, \emph{i.e.}, the number of cells whose state constitutes an input for the function $f$. Eq.~(\ref{extended})
 corresponds to $r=3$.
The case in which the function $f$ is symmetric with respect to all inputs defines \emph{totalistic} CA, since in this case one can consider that the value of the function $f$ depends only on the sum of the values of sites in the neighborhood. While generic CA with range $r$ are defined by $2^r$ entries in the look-up table, totalistic CA are defined by $r+1$ entries. By arranging the output values of the look-up table as Boolean digits, one can compactly represent a CA rule as an integer number $R$, as shown for instance in Figure~\ref{fig:10}. 

Cellular automata may exhibit a large variety of dynamical behaviors. The number of possible states of a lattice of $L$ Boolean cells is finite, and equal to $2^L$. Since the dynamics is deterministic, only limit cycles attractors are possible. One can divide the possible scenarios according with the number of attractors, the distribution of their basins and their period. For instance with $r=3$, trivial rules like rule 0 have only one attractor with a large basin and period equal to one. The identity rule (which is not totalistic) 
has a large number of attractors ($2^L$), each one with one state 
and period 1. The majority rule $1100|_2=12|_{10}$ has a intermediate 
number of attractors with short periods (1). ``Chaotic'' rules like rule 
$1010|_2=10|_{10}$ exhibit cycles with very long period of the order of 
the total number of configurations as in Figure~\ref{fig:pattern}. 
Since in this case the period scales as an exponential of the size of the system, the difference between a periodic and aperiodic trajectory is not relevant (statistical quantities take similar values). Moreover, a defect or damage typically spreads in the configuration.

It is possible~\cite{BoolDeriv} to extend the concept of derivative to cellular automata The Boolean derivative of $\v{F}$ is the Jacobian matrix with
components
\begin{equation}
\begin{aligned}
  J_{i,j}=&\dfrac{\partial F_i(\v{x})}{\partial x_j}
  =F_i(x_0,\dots,x_j\oplus 1,\dots,x_{N-1})\oplus F_i(x_0,\dots,x_j,\dots,x_{N-1})\\
  =&\begin{cases}
    1 & F_i\: \text{changes when}\: x_j\: \text{changes},\\
    0 & F_i\: \text{does not change when}\: x_j\: \text{changes},
    \end{cases}\label{BooleanJacobian}
\end{aligned}
\end{equation}
where $\oplus$ denotes the sum modulo two.

\begin{figure}
  \begin{center}
   \begin{tabular}{cc}
    damage spreading & evolution in tangent space \\[3mm]
    {\small $\v{u}^{(t+1)}=\v{F}(\v{x}^{(t)})\oplus \v{F}(\v{x}^{(t)}\oplus \v{u}^{(t)})$}&%
    {\small $\v{u}^{(t+1)}=\v{J}(\v{x}^{(t)})\cdot\v{u}^{(t)}$}\\[3mm]
    \includegraphics[width=0.4\textwidth]{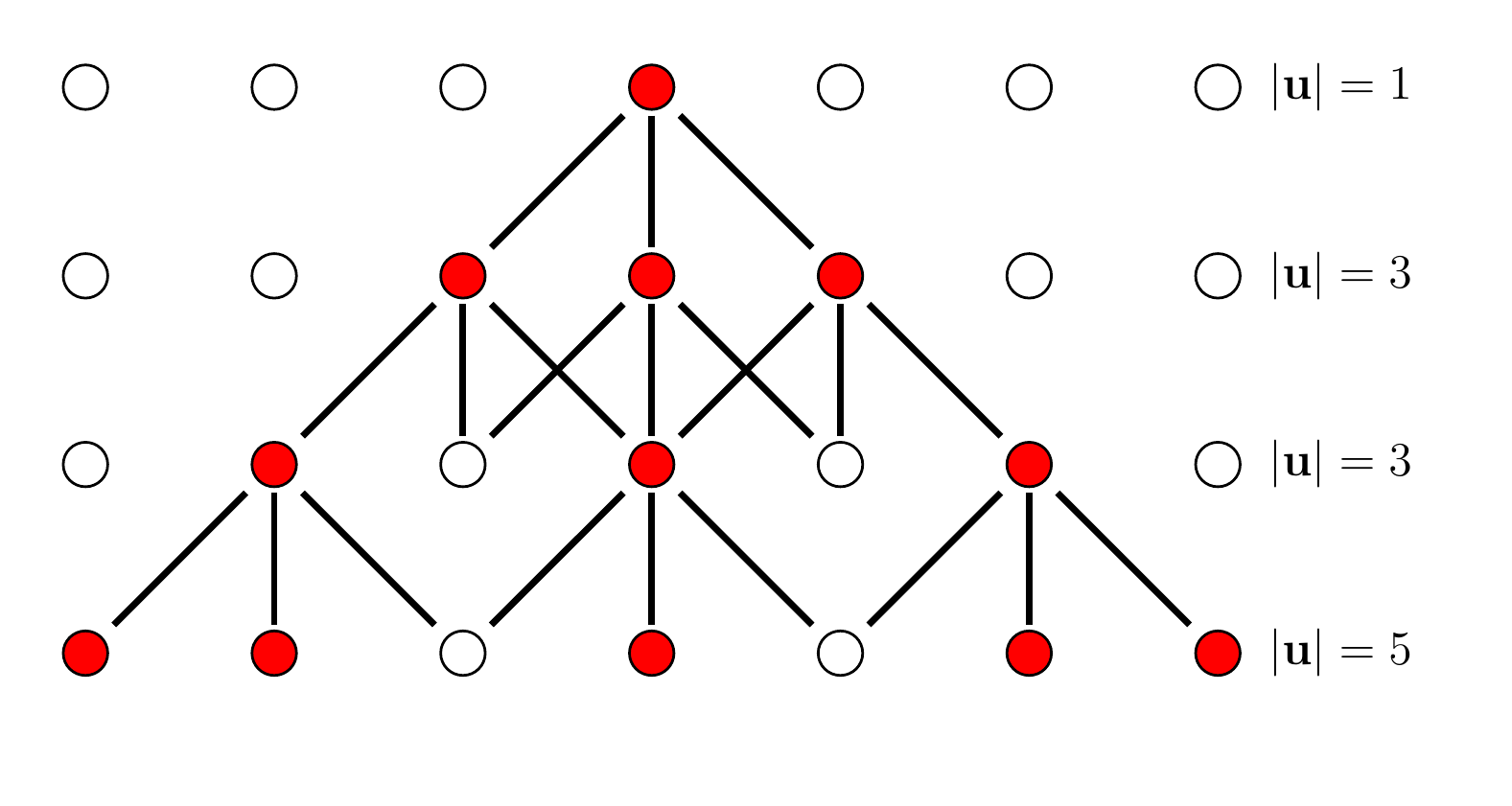} &
    \includegraphics[width=0.4\textwidth]{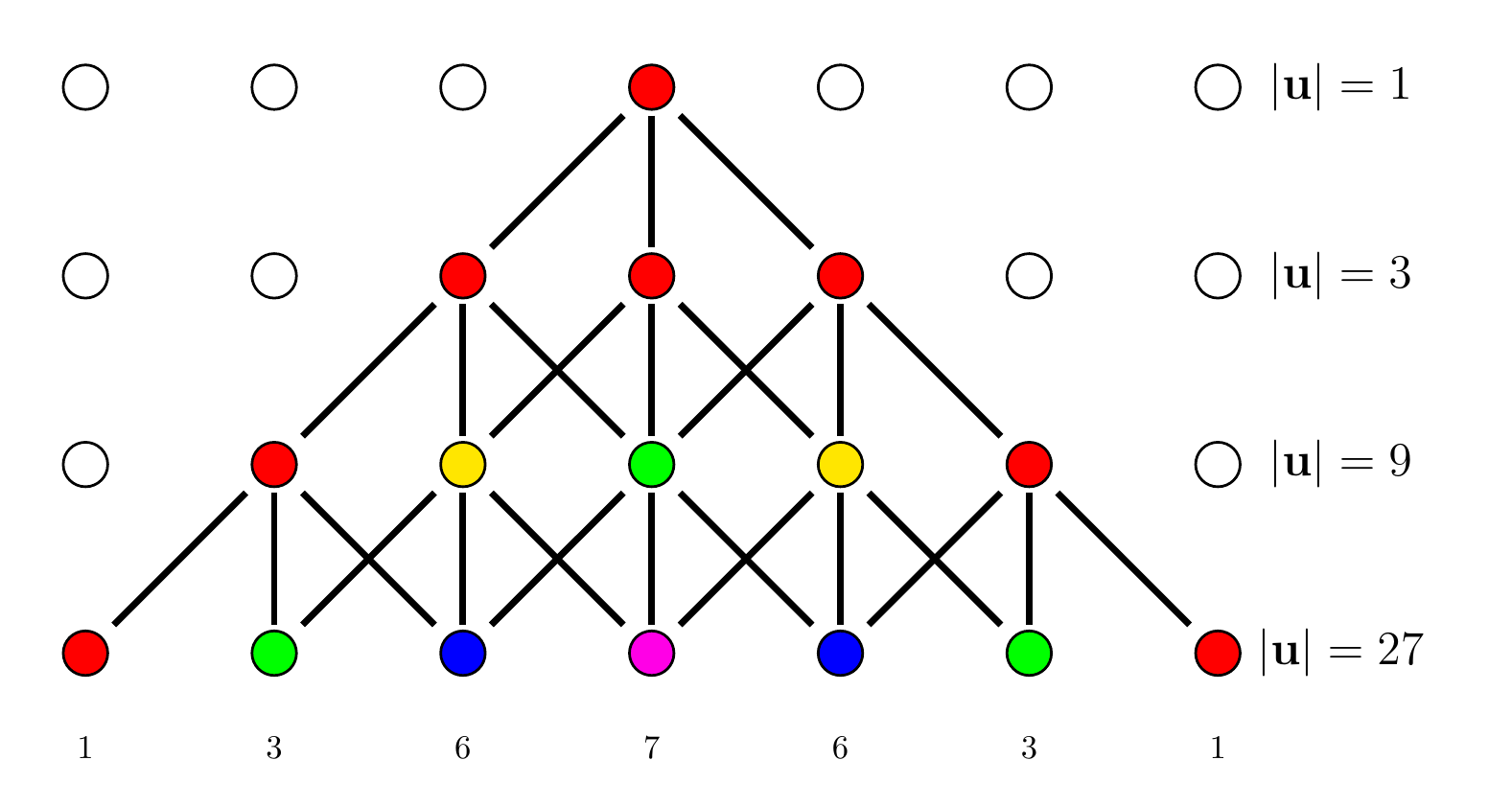}
   \end{tabular}
   \end{center}
  \caption{\label{fig:damage} Difference between damage spreading and evolution in tangent space}
\end{figure}

Many ``standard'' results may be extended to Boolean
derivatives, for instance the Taylor expansion
\[
  f(x,y) = \left(\frac{\partial f}{\partial
  x}\right)_{{x=y=0}} x \oplus
  \left(\frac{\partial f}{\partial y}\right)_{{x=y=0}} y \oplus
  \left(\frac{\partial^2 f}{\partial x\partial
  y}\right)_{{x=y=0}} xy.
\]
One can apply the ``linear development'' to discrete damages, and define  a 
discrete Jacobian matrix, Eq.~\eqref{BooleanJacobian}. Similarly to 
continuous systems, it is possible to define the largest Lyapunov 
exponent~\cite{CASynchro} related to the synchronization threshold 
(see Section~\ref{sec:synchro}). In contrast to continuous dynamics, 
defects can self-annihilate, so that the actual development of damage is 
different from the linearized one  and they coincide only in the limit of 
vanishing damage, as shown in Figure~\ref{fig:damage}.

\subsection{Synchronization of extended systems}\label{sec:synchro}

There are many ways of ``pushing'' together two extended replicas. One possibility is ``uniform'' pushing
\[
   y_i(t+1) = (1-p) F_i (\v{y}(t)) + p F_i (\v{x}(t)),
\]
for which the analysis presented above applies, with $ p_c = 1-\exp(-\lambda_0)$. This control is however quite difficult to be implemented experimentally in an extended system.
Uniform synchronization of chaotic maps gives results similar to low-dimensional systems: $p_c=1-\exp(-\lambda_0)$. 

Another possibility is that of ``pinching'' synchronization
\[
   y_i(t+1) =   \begin{cases}
                   F_i (\v{y}(t)) & \text{with probability $1-p$},\\
                   F_i (\v{x}(t)) & \text{with probability $p$}.\\
                  \end{cases}
\]
In pinching synchronization, one has the possibility of applying the synchronization ``strength'' to a suitably chosen subset of sites. 
Pinching synchronization depends on coupling: uncoupled chaotic maps
synchronizes for $p_c=0$. In general $p_c$ is larger for larger
couplings~\cite{StableChaos1}.
 
In synchronization problems, synchronization is applied ``blindly''. In control problems, the goal is that of exploiting available information in order
to apply a smaller amount of control (or achieve a stronger synchronization).

\section{Control of CA} \label{sec:CAcontrol}

We study here the spatial application of synchronization
\[
  \left\{\begin{aligned}
    \v{x}' &= \v{F}(\v{x}),\\
    \v{y}' &= (\v{1}-\v{p})\odot \v{F}(\v{y}) \oplus \v{p}\odot \v{F}(\v{x}),
  \end{aligned}\right.
\]
where $\odot$ is the Hadamard (component by component) product, and the effect of synchronization $p_i\in \{0,1\}$ may depend on the position $i$.
Therefore, the difference $\v{u}$ evolves as 
\begin{equation}
    \v{u}' = (\v{1}-\v{p})\odot \bigl(\v{f}(\v{x})-\v{f}(\v{y})\bigr)\label{u}
\end{equation}
and in the limit of vanishing distance,
 \begin{equation}
    \v{u}'\simeq (\v{1}-\v{p})\odot \v{J} \v{u},\label{ulin}
\end{equation}
where the scalar product $\v{J} \v{u}$ is computer modulo two.
 The control parameter is the average synchronization effort $k=(\sum_i p_i)/N$.
The efficacy of synchronization (order parameter) is the asymptotic distance $h=(\sum_i u_i)/N$.

It is possible in principle to find the absolute minimum of $k$ by computing the effects of all possible choices of $p_i$, given an initial configuration $\v{x}_0=\v{x}(0)$. This constitutes a great computational load.
Since we are interested in possible real-time applications, we impose that the choice of $p_i=1$ may only depend on \emph{local} information: the neighborhood configuration and a $t=1$ time window.

\subsection{Implementation of control}

We investigate the following cases concerning how control $\v{p}$ is applied:
\begin{enumerate}
  \item Blindly with probability $p=k$ (standard pinching synchronization).
  \item With a probability $p$ proportional to the sum of the first-order derivatives.
  \item With a probability $p$ inversely proportional to the sum of first-order derivatives.
\end{enumerate}
In order to keep the implementation simple, instead of fixing $k$ and computing the probability $p$, we let $p$ be a free parameter, and measure the actual fraction of synchronized sites $k$ and the average asymptotic distance $h$.
The previous schemes only require information about $\v{x}$. If information about $\v{y}$ or about the damage distribution $\v{u}$ is available, the cost $k$ is reduced by a factor $h$, since in this case we can apply the rule only when it is needed.

\begin{figure}
  \begin{center}
  \begin{tabular}{ccc} 
    \includegraphics[width=0.3\columnwidth]{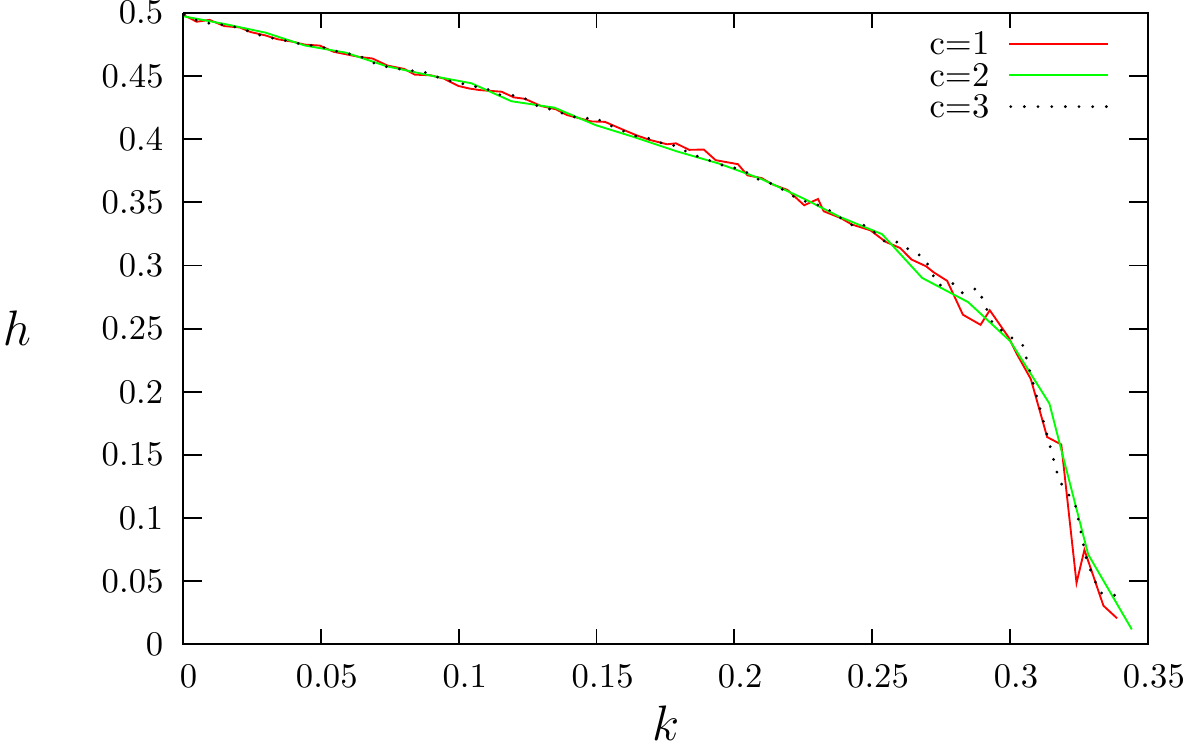} &
    \includegraphics[width=0.3\columnwidth]{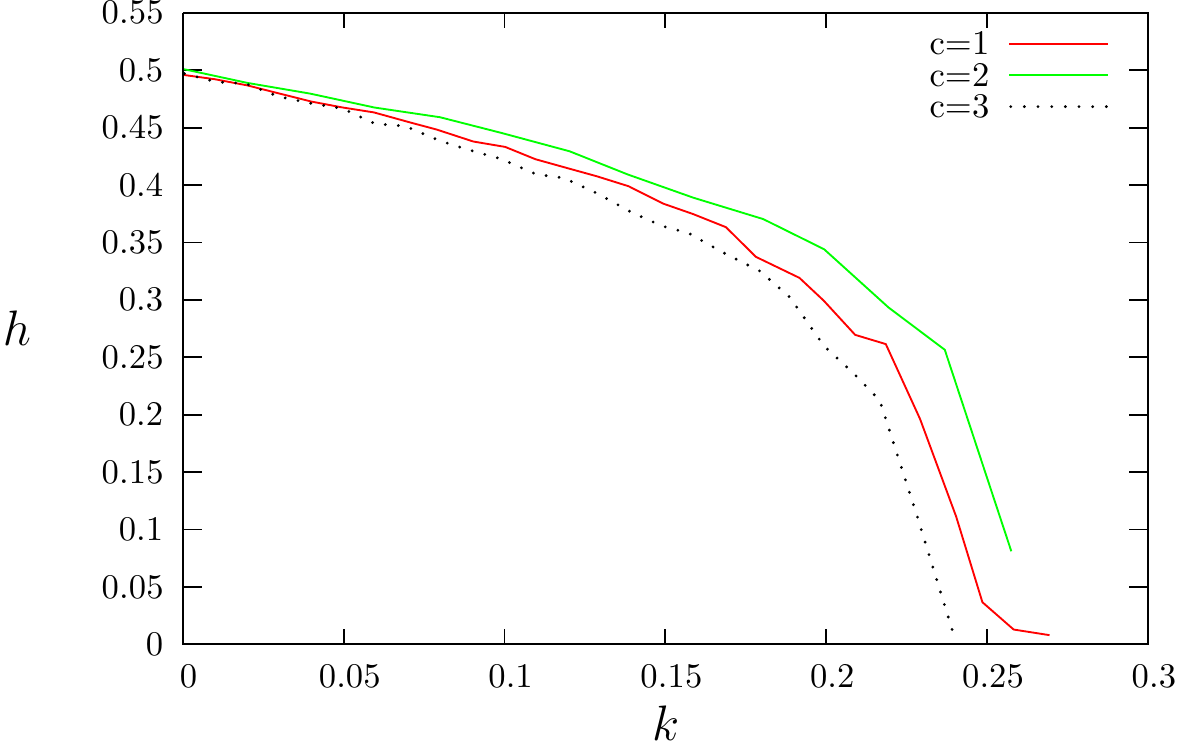}&
    \includegraphics[width=0.3\columnwidth]{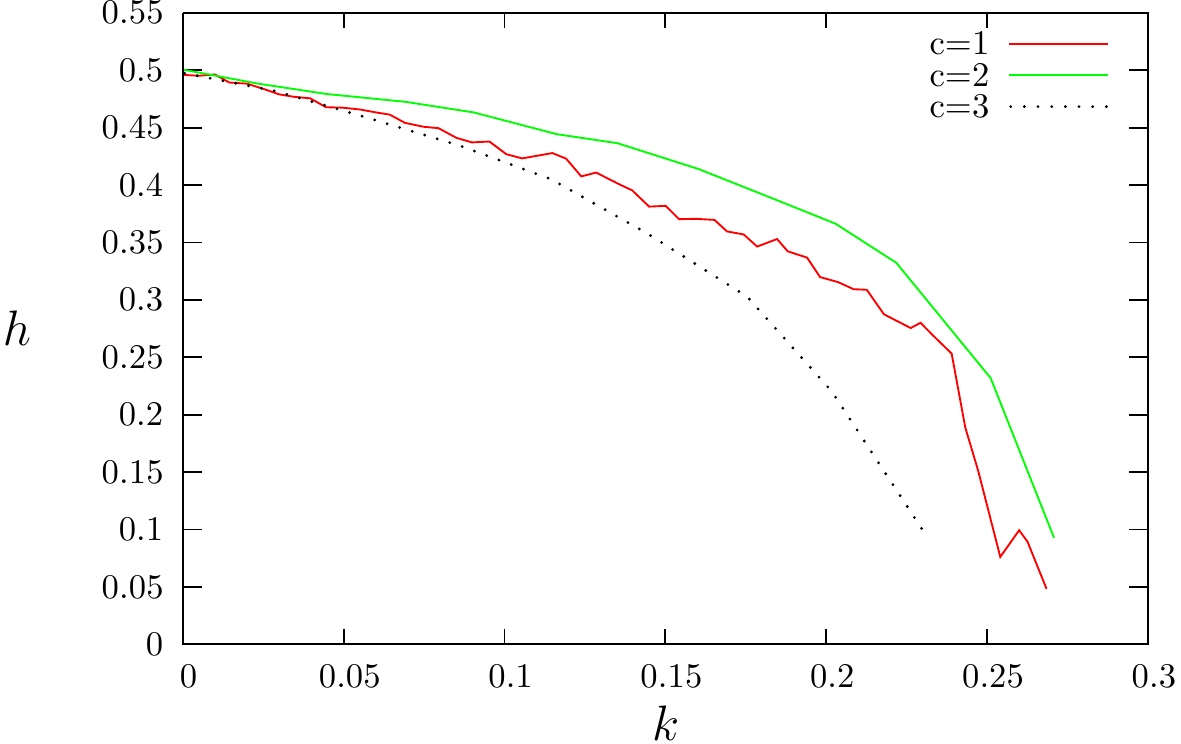}\\
    (a) & (b) & (c)
  \end{tabular}
  \end{center}
 \caption{\label{fig:control} Plots of the different types of control for: (a) $r=3$,  $R=1$ (linear rule);
 (b) $r=3$, $R=6$ (nonlinear rule); (c) $r=6$, $R=30$ (nonlinear rule). 
 For nonlinear rules, control 2 is worse and control 3 is better than blind one (control type 1).}
\end{figure}

\begin{figure}
  \begin{center}
  \begin{tabular}{ccc}
  \includegraphics[width=0.3\columnwidth]{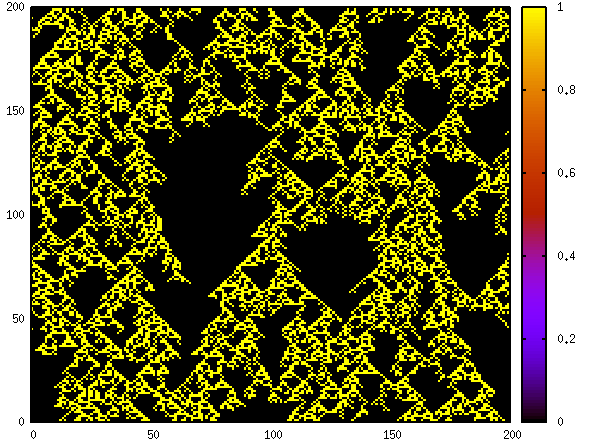}&
  \includegraphics[width=0.3\columnwidth]{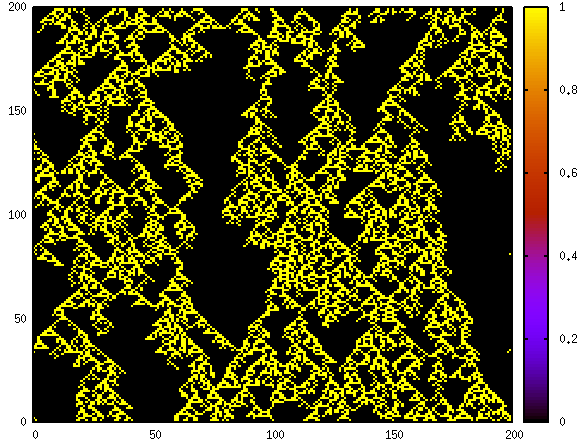}&
  \includegraphics[width=0.3\columnwidth]{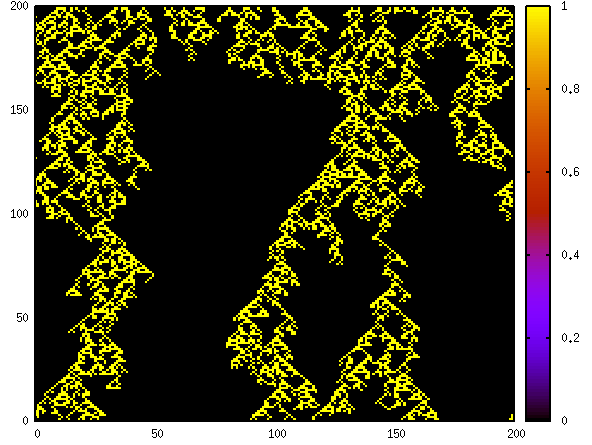}\\
  control 1 & control 2 & control 3\\
  \end{tabular}
  \end{center}
 \caption{\label{selfannihilation} 
 Time evolution of defects for different types of control. Here $r=3$ and $R=6$, all cases starting from the same configuration. The effective probability $p$ has been chosen so to have the same average control $k$ in the three cases. One can notice that clusters of defects for control 3 are less dense than that of control 1 and 2.}
\end{figure}

\subsection{Results}

Preliminary simulation results are presented in Figure~\ref{fig:control}.
As expected, for linear rules there is no influence of the type of control, 
since all configurations have the same number of derivatives. For nonlinear rules, the observed behavior is the opposite of what is expected for continuous systems. Control 2, that minimizes the distance $h$ for vanishing number of damages according to Eq.~(\ref{ulin}), gives worse results than the blind control 1. Control 3,
inversely proportional to the sum of first-order derivatives, gives better results than the blind control 1.
This result holds also for larger neighborhoods (Figure~\ref{fig:control}-c), but not for all rules.
   
This surprising effect may be due to the fact that defects self-annihilate, as shown in Figure~\ref{selfannihilation}. In other words, we can exploit the characteristics of cellular automata (and other stable chaotic systems) in order to achieve a better control by exploiting the local contraction of the evolution rule. 
  
\section{Conclusions}

Spatially extended stable systems (namely cellular automata) may exhibit unpredictable behavior (finite-distance chaoticity). The pinching synchronization threshold is related to this chaoticity. On the other hand, Boolean derivatives and discrete Lyapunov exponents may be used to characterize this kind of chaos.
Synchronization may also be exploited for control in experimental situations. In the control problem one aims at discovering a protocol that keeps
the distance $h$ below a certain threshold with the minimum ``effort'', given some constraints. We have chosen to investigate the behavior of two control schemes based on the local number of non-zero first-order derivatives, taking as reference the ``blind'' pinching synchronization protocol.

We have shown that, differently from usual chaotic systems, one can exploit \emph{self-annihilation} of defects to
obtain synchronization with a weaker  control, corresponding to the case in which the control is inversely proportional to the number of non-zero derivatives.
 
\section*{Acknowledgments}

Partial economic support from CONACyT project 25116 is acknowledged.

\end{document}